\RequirePackage{fix-cm}
\documentclass[twocolumn,epjc3]{svjour3}  
\smartqed  
\RequirePackage{graphicx}

\usepackage{atlasphysics}
\usepackage{lineno}
\usepackage{xspace}
\usepackage{slashbox}
\usepackage{multirow}
\usepackage{cite}
\usepackage{array}
\usepackage[hyperindex,breaklinks]{hyperref} 

\usepackage{preprintcover}

\newcommand{\threebTitle}{Search for top and bottom squarks from gluino pair production in final states with 
missing transverse energy and at least three $b$-jets with the ATLAS detector}

\PreprintCoverPaperTitle{\threebTitle}

\PreprintIdNumber{CERN-PH-EP-2012-194}  
\PreprintCoverAbstract{This letter reports the results of a search for top and bottom squarks from gluino pair production 
in \totallumi~of $pp$ collisions at $\sqrt{s}=7$~\TeV\ using the ATLAS detector at the LHC. The search is performed in events with large 
missing transverse momentum and at least three jets identified as originating from a $b$-quark. Exclusion limits are presented for a variety 
of gluino-mediated models with gluino masses up to 1~\TeV\ excluded.}
\PreprintJournalName{European Physical Journal C}

\newcommand{\CRA}   {CR4-60}
\newcommand{\CRB}   {CR4-70}
\newcommand{\CRC}   {CR6-70}
\newcommand{\CRD}   {CR6-75}

\newcommand{\SRA}   {SR4-L}
\newcommand{\SRB}   {SR4-M}
\newcommand{\SRC}   {SR4-T}
\newcommand{\SRD}   {SR6-L}
\newcommand{\SRE}   {SR6-T}

\newcommand{\sqR}   {\tilde{q}^{}_{R}}
\newcommand{\sqL}   {\tilde{q}^{}_{L}}

\newcommand{\bone}   {\tilde{b}^{}_{1}}
\newcommand{\tone}   {\tilde{t}^{}_{1}}
\newcommand{\neut}  {\tilde{\chi}^{0}_{1}}  
\newcommand{\chipm}  {\tilde{\chi}^{\pm}_{1}} 
\def\etmiss{\ensuremath{E_{\mathrm{T}}^{\mathrm{miss}}}\xspace}

\def\dphimin{\ensuremath{\mathrm{\Delta}\phi_{\mathrm{min}}}}

\def\bjets{$b$-jets\xspace}

\def\btagging{$b$-tagging\xspace}
\def\btag{\ensuremath{b}-tag\xspace}

\def\bquarks{$b$-quarks\xspace}
\def\meff{\ensuremath{m_{\mathrm{eff}}}\xspace}

\newcommand{\gl}   {\tilde{g}}
\newcommand{\sq}   {\tilde{q}}

\renewcommand{\ttbar} {\ensuremath{t\bar{t}}\xspace}

\renewcommand{\met} {\ensuremath{E_{\mathrm{T}}^{\mathrm{miss}}}\xspace}
\renewcommand{\pt} {\ensuremath{p_\mathrm{T}}\xspace}

\newcommand{\lumi}{4.7}
\newcommand\totallumi{4.7~${\rm fb}^{-1}$}

\def\met{\ensuremath{E_{\mathrm{T}}^{\mathrm{miss}}}}

\newcommand{\mywidth}{0.9}

\journalname{Eur. Phys. J. C}

\begin{document}

\title{\threebTitle}
\titlerunning{Search for top and bottom squarks in final states with 
missing transverse energy and at least three $b$-jets}
\author{The ATLAS Collaboration}
\date{July 19, 2012}

\maketitle
\normalsize

\begin{abstract} 
This letter reports the results of a search for top and bottom squarks from gluino pair production in \totallumi~of $pp$ collisions 
at $\sqrt{s}=7$~\TeV\ using the ATLAS detector at the LHC. The search is performed in events with large missing transverse momentum 
and at least three jets identified as originating from a $b$-quark. Exclusion limits are presented for a variety of gluino-mediated models 
with gluino masses up to 1~\TeV\ excluded.
\end{abstract}

Supersymmetry (SUSY)~\cite{Miyazawa:1966,Ramond:1971gb,Golfand:1971iw,Neveu:1971rx,Neveu:1971iv,Gervais:1971ji,Volkov:1973ix,Wess:1973kz,Wess:1974tw} 
provides an extension of the Standard Model (SM) which resolves the hierarchy problem \cite{Weinberg:1975gm,Gildener:1976ai,Weinberg:1979bn,Susskind:1978ms} 
by introducing supersymmetric partners of  the known bosons and fermions. In the framework of the $R$-parity conserving minimal supersymmetric extension 
of the SM (MSSM)~\cite{Fayet:1976et,Fayet:1977yc,Farrar:1978xj,Fayet:1979sa,Dimopoulos:1981zb}, SUSY particles are produced in pairs and the lightest 
supersymmetric particle (LSP) is stable, providing a possible candidate for dark matter. In a large variety of models, the LSP is the lightest 
neutralino ($\neut$). The coloured superpartners of quarks and gluons, the squarks ($\sq$) and gluinos ($\gl$), if not too heavy, would be produced 
in strong interaction processes at the Large Hadron Collider (LHC) and decay via cascades ending with the LSP. The undetected LSP results in missing 
transverse momentum -- whose magnitude is referred to as \met\ -- while the rest of the cascade yields final states with  multiple jets and possibly leptons. 
In the MSSM, the right-handed and left-handed squarks, $\sqR$ and $\sqL$, can mix to form two mass eigenstates $\tilde{q}_1$ and $\tilde{q}_2$. 
The mixing effect is proportional to the masses of the SM fermion partners and can therefore be large for the third generation. This may lead to 
the lightest sbottom ($\bone$) and stop ($\tone$) mass eigenstates being much lighter than the other squarks. As a consequence, $\bone$ and $\tone$ 
could be produced with relatively large cross-sections at the LHC, either directly in pairs, or through $\gl\gl$ production followed by 
$\gl \rightarrow \bone b$ or $\gl \to \tone t$ decays. 

This letter extends the search for gluino-mediated $\bone$ and $\tone$ production at ATLAS reported earlier~\cite{PRD}, 
which used 2.05~fb$^{-1}$ of data collected in the first half of 2011 at a centre-of-mass energy of 7~TeV. 
The present analysis comprises the full 2011 dataset of  \lumi~$\ifb$ and adopts an improved selection that 
requires large \met, no electron or muon and at least three jets identified as originating from 
$b$-quarks ($b$-jets) in the final state. Results are interpreted in four simplified models where 
sbottoms or stops are the only squarks produced in the gluino decays, leading to final states with four \bquarks. 
Searches in similar scenarios have also been reported by the CMS Collaboration~\cite{cms_2llSS}. 

The ATLAS detector~\cite{DetectorPaper:2008} consists of inner tracking devices 
surrounded by a superconducting solenoid, electromagnetic and hadronic calorimeters and a muon spectrometer 
with a toroidal magnetic field. The inner detector provides precision tracking of charged particles for  
$|\eta| < 2.5$\footnote{ATLAS uses a right-handed coordinate system with its origin at the nominal interaction point (IP) in the centre of 
the detector and the $z$-axis along the beam pipe. The $x$-axis points from the IP to the centre of the LHC ring, and the $y$ axis points upward. 
Cylindrical coordinates $(r,\phi)$ are used in the transverse plane, $\phi$ being the azimuthal angle around the 
beam pipe. The pseudorapidity is defined in terms of the polar angle $\theta$ as $\eta=-\ln\tan(\theta/2)$.The 
distance~$\mathrm{\Delta} R$ in the $\eta-\phi$ space is defined as $\mathrm{\Delta} R = \sqrt{(\mathrm{\Delta}\eta)^2+(\mathrm{\Delta}\phi)^2}$.}. 
It is immersed in a $2$\,T magnetic field from the solenoid and consists of a silicon pixel detector, a silicon strip 
detector and a straw tube tracker that also provides transition radiation measurements for electron identification. 
The calorimeter system covers the pseudorapidity range $|\eta| < 4.9$. It is composed of sampling calorimeters with 
either liquid argon (LAr) or scintillating tiles as the active medium. The muon spectrometer has separate trigger and high-precision 
tracking chambers which provide muon identification and momentum measurement for $|\eta| < 2.7$. 

Samples of simulated events are used for the description of the background and 
to model the SUSY signal. The dominant sources of background come from events with 
$b$-quarks in the final state. Monte Carlo (MC)  samples of \ttbar, $W$/$Z$ and diboson  events  
produced in association with light- and heavy-flavour jets are generated with 
{\tt ALPGEN}~\cite{alpgen} and the parton distribution function (PDF) set {\tt CTEQ6L1}~\cite{cteq6l1}.  
These samples are generated with different maximum numbers of additional partons at the matrix-element level. 
Diboson samples are generated with up to three additional partons, $\ttbar$+jet and $Z$($\rightarrow \ell^+ \ell^-$)+jets  
($\ell$ = $e$,$\mu$,$\tau$) samples with up to five additional partons, and $W$($\rightarrow \ell \nu$)+jets and 
$Z$($\rightarrow \bar{\nu} \nu$)+jet samples with up to six additional partons. Single top quark production is 
simulated with {\tt MC@NLO}~\cite{mcatnlo} using the next-to-leading-order (NLO) PDF set {\tt CTEQ6.6}~\cite{cteq66}. 
The fragmentation and hadronisation for the {\tt ALPGEN} and {\tt MC@NLO} samples are performed with 
{\tt HERWIG}~\cite{herwig,herwig65}, using {\tt JIMMY}~\cite{jimmy} for the underlying event. Samples of 
$\ttbar$+$W$, $\ttbar$+$Z$ and $\ttbar$+$WW$ events are generated with {\tt MADGRAPH}~\cite{madgraph} 
interfaced to {\tt PYTHIA}~\cite{pythia}. The signal samples are generated using {\tt Herwig++}~\cite{Bahr:2008pv}. 
The MC samples are processed through the ATLAS detector simulation~\cite{atlassimulation} based on {\tt GEANT4}~\cite{geant4} 
taking into account the effect of multiple $pp$ interactions per bunch crossing. For the comparison with data, all SM background 
cross-sections are normalised to the results of higher-order calculations when available, using the same values as in Ref.~\cite{PRD}. 

Jets are reconstructed from three-dimensional calorimeter energy clusters using the anti-$k_t$ jet algorithm~\cite{Cacciari:2008gp,Cacciari200657} 
with a radius parameter of 0.4. The measured jet energy is corrected for inhomogeneities and for the non-compensating nature of the calorimeter 
by using $\pt$- and $\eta$-dependent correction factors, and additional corrections for pile-up are applied~\cite{JES}. 
Jets are required to have $\pT > 20$~\GeV, and are reconstructed in the range  $|\eta| < 4.9$. 
Events are rejected if they include jets failing the quality criteria described in Ref.~\cite{JES}, or if there is  
any selected jet with $|\eta| < 2$ for which the scalar sum of the transverse momenta of its associated tracks 
 is less than 5\% of the jet $\pt$. A neural-network-based algorithm~\cite{btag1} is used to identify jets containing a $b$-hadron decay. 
This uses as inputs the output weights of different algorithms exploiting the   impact parameter 
of the inner detector tracks, the secondary vertex reconstruction and the topology of $b$- and $c$-hadron decays inside the jet. 
Three operating points are used, corresponding to efficiencies of 60\%, 70\% and 75\% for tagging \bjets in a MC sample of \ttbar\  
events. In all cases the tagging rate is less than 2\% for light-quark and gluon jets, 10\% for $\tau$ leptons decaying hadronically 
and 25\% for $c$-quark jets. The $b$-jets are identified within the nominal acceptance of the inner detector ($|\eta| < 2.5$) and are 
required to have $\pt > 30$~\GeV. To compensate for the differences between the $b$-tagging efficiency and the mistag rates in data 
and MC simulation, $b$-tagging scale factors are applied to each jet in the simulations, as described in Refs.~\cite{btag1,btag2,btag3}. 

Electrons are reconstructed from energy clusters in the electromagnetic calorimeter matched to a track in the inner detector. Electron 
candidates are required to have $\pt>20$~\GeV\ and $|\eta|<2.47$ and must satisfy the  ``medium'' selection criteria described in 
Ref.~\cite{2011mk}. Muons candidates are identified using a match between an extrapolated inner detector track and one or more track 
segments in the muon spectrometer, and are required to have $\pt>10$~\GeV\ and $|\eta| < 2.4$. 

To resolve overlaps between reconstructed jets and leptons, jets within a distance of $\mathrm{\Delta} R=0.2$ of an electron candidate are 
rejected. Furthermore, any lepton candidate  with a distance  $\mathrm{\Delta} R < 0.4$  to the closest remaining jet is 
discarded. Events containing any remaining electrons and muons  are vetoed in the control and signal regions defined in Tables 1 and 2. 

The measurement of the missing transverse momentum two-dimensional vector (and its magnitude \etmiss) is based on the transverse momenta 
of all remaining jets with $|\eta|<4.9$, all electron and muon candidates and all calorimeter clusters not associated to such objects. 

Events are selected using triggers requiring one high \pt jet and \met. Different trigger thresholds were used to cope with the increasing 
luminosity. These triggers are fully efficient for this analysis, which requires one jet with $\pt>130$~\GeV\ and $\met>160$~\GeV\ at the 
offline reconstruction stage. Events must pass basic quality criteria to reject detector noise and non-collision backgrounds. They are also 
required to have a reconstructed primary vertex associated with five or more tracks with $\pt >0.4$~\GeV; when more than one such vertex is 
found, the vertex with the largest summed $\pt^2$ of the associated tracks is chosen as the hard interaction vertex. Events are required to 
have at least three $b$-tagged jets, and two jet-multiplicity regions ($N_J \geq 4$ and $N_J \geq 6$) are considered by selecting 
jets with $|\eta|<2.8$ and $\pt>50$~\GeV.  

Two variables are calculated from the reconstructed objects to further select the events: \meff and \dphimin. The effective mass \meff is 
defined as the scalar sum of the \met\ and the $\pt$ of all selected jets in a given jet-multiplicity region. The  \dphimin\ is defined as 
the minimum azimuthal separation between the selected jets and the missing transverse momentum direction. Placing the requirements  
$\dphimin>0.4$ and $\met/\meff > 0.2$ reduces the amount of multi-jet background, where $\met$ results from mis-reconstructed jets or from 
neutrinos emitted close to the direction of the jet axis. 

Two sets of signal regions are defined which yield good signal sensitivity for the various models and parameter values studied here. 
They are characterised by having at least four (SR4) or six (SR6) jet candidates, no electron or muon, and  are further classified 
as loose (L), medium (M) or tight (T) depending on the $\met$ and  $\meff$ thresholds and on the \btagging operating point. The requirements 
that characterise each signal region are summarised in Table~\ref{tab:SR}. 

\begin{table}
\centering
\begin{tabular}{c|c|c|c|c}
\hline 
\multicolumn{5}{c}{Common criteria: lepton veto,  $\pt^{j_1} > 130$ \gev,} \\
\multicolumn{5}{c}{ $\geq$ 3 $b$-jets,  $\met / \meff> 0.2$, $\dphimin>0.4$} \\
\hline 
SR  & $N_J$ & $\met$ & $\meff$ & \btag OP  \\ \hline 
SR4-L & $\geq$ 4j &  $>$160 \gev\  &  $>$500 \gev\ & 60\%  \\
SR4-M & $\geq$ 4j &  $>$160 \gev\  &  $>$700 \gev\ & 60\%  \\
SR4-T & $\geq$ 4j &  $>$160 \gev\  &  $>$900 \gev\ & 70\%  \\
\hline 
SR6-L & $\geq$ 6j &  $>$160 \gev\  &  $>$700 \gev\ & 70\%  \\
SR6-T & $\geq$ 6j &  $>$200 \gev\  &  $>$900 \gev\ & 75\%  \\
\hline
\end{tabular}
\caption{Definition of the five signal regions based on the number of jets ($N_J$), the 
$\met$, $\meff$ requirements and the $b$-tagging operating point (OP).} 
\label{tab:SR}                
\end{table}

The main source of reducible background is the production of \ttbar\ events in association with additional jets followed by the leptonic decay 
of one $W$ boson, where the lepton is not reconstructed or is misidentified as a jet (mainly through the hadronic decays of a $\tau$ lepton). 
This background is estimated by normalising the MC event yield in the signal region to the extrapolated event yield observed in a \ttbar-dominated 
control region. Systematic uncertainties that are correlated between the control and the signal regions largely cancel out in this procedure. 
Additional sources of reducible background are single top, $\ttbar$+$W$/$Z$ and  $W$/$Z$+heavy-flavour jets. Their contributions are taken from 
MC simulation and account for 10\% to 20\% of the total background depending on the signal region. The irreducible background $\ttbar$+$b\bar{b}$ 
is also estimated from MC simulation and accounts for about 10\% of the total background in all signal regions. The reducible contribution from 
multi-jet events is estimated with a data-driven method, based on a jet response smearing technique~\cite{jet_smearing}, and is found to account 
for less than 5\% of the total background in all signal regions. 

Four control regions where the $\ttbar$+jets background accounts for more than 80\% of the total yield are defined by applying the same jet 
requirements and lepton veto as in the signal regions, but requiring exactly two $b$-jets instead of three or more. The requirements 
$\meff>500$~\GeV\ and $\met>160$~\GeV\ are applied to  all control regions to make them kinematically similar to the signal regions, while 
reducing the contamination from possible SUSY signal events. The definition of the control regions  is summarised in  Table~\ref{tab:CR}. 
The numbers of expected SM events in the four control regions, as predicted by the jet smearing technique for multi-jet events and by MC simulation 
for other processes, are compared to those observed in data in Table~\ref{tab:CR_results}. The results agree within experimental errors. 

\begin{table}
\centering
\begin{tabular}{c|c|c|c}
\hline 
\multicolumn{4}{c}{Common criteria: lepton veto, $\pt^{j_1} > 130$ \gev,}\\ 
\multicolumn{4}{c}{ = 2 \bjets, $\met / \meff$ $> 0.2$, $\dphimin > 0.4$,} \\ 
\multicolumn{4}{c}{ $\met > 160$ \GeV, $\meff > 500$ \GeV } \\
\hline 
CR  & $N_J$ & \btag OP  & corresponding SR \\ \hline 
\CRA & $\geq$ 4j &  60$\%$  & \SRA, \SRB \\
\CRB & $\geq$ 4j & 70$\%$ & \SRC \\
\hline 
\CRC & $\geq$ 6j  &  70$\%$  & \SRD \\
\CRD & $\geq$ 6j &  75$\%$ &   \SRE \\
\hline 
\end{tabular}
\caption{Definition of the four control regions used to estimate the $\ttbar$+jets background.} 
\label{tab:CR}                
\end{table}

\begin{table}[htp]
\centering
\begin{tabular}{c|c|c|c|c}
\hline
CR & $\ttbar$+jets &  others  & SM & data  \\
\hline
\CRA & 330 $\pm$ 90 &  65 $\pm$ 25 & 395 $\pm$ 115 & 402  \\
\hline
\CRB & 490 $\pm$ 125 & 100 $\pm$ 35 & 590 $\pm$ 160 & 515 \\
\hline
\CRC & 38 $\pm$ 11 &  7 $\pm$ 3 & 45 $\pm$ 13 & 46 \\
\hline
\CRD & 40 $\pm$ 12 & 10 $\pm$ 4 & 50 $\pm$ 15 & 52 \\
\hline
\end{tabular}
\caption{Expected numbers of SM events and observed data events in the four \ttbar\ control regions. 
The contribution from  $\ttbar$+jets events is taken directly from MC simulation. The column ``others'' includes 
the contributions from single top,  $\ttbar$+$b\bar{b}$, $\ttbar$+$W$/$Z$ and  $W$/$Z$+jets processes, also estimated 
from MC simulation, and the contribution from multi-jet events which is estimated with the jet smearing technique and  
accounts for less than 5\% of the total background. The column ``SM'' shows the total expected background and is the sum 
of the columns ``$\ttbar$+jets'' and ``others''. The uncertainties include all detector-related systematic uncertainties.} 
\label{tab:CR_results}
\end{table}

The dominant detector-related systematic effects are due to the jet energy scale (JES) and resolution (JER) uncertainties, and 
the uncertainty on the $b$-tagging efficiency and mistag rates. The JES uncertainty is derived from a combination of simulations, 
test beam data and in-situ measurements~\cite{JES}, and includes additional uncertainties due to the jet flavour and nearby jets. 
Uncertainties on the JER are obtained with an in-situ measurement of the jet response asymmetry in di-jet events. These uncertainties  
on jets are propagated to the \met\ measurement, and additional uncertainties on \met\  arising from energy deposits not associated with 
any reconstructed objects are also included. The $b$-tagging uncertainty is evaluated by varying the $\eta$-, $\pt$- and flavour-dependent 
scale factors applied to each jet in the simulation within a range that reflects the systematic uncertainty on the measured tagging efficiency 
and mistag rates.

The systematic uncertainties in the modelling of the $\ttbar$+jets background are assessed as follows: the uncertainty due to the choice of 
the MC generator is estimated by comparing the leading-order {\tt ALPGEN} generator to the MC@NLO generator; the uncertainty due to the 
factorisation and matching scale ambiguities in {\tt ALPGEN} are estimated by independently varying their nominal settings by factors of one 
half and two; the uncertainty due to the finite number of additional partons at the matrix-element level is assessed by comparing 
inclusive {\tt ALPGEN} samples generated with up to three and up to five extra partons. Finally the PDF uncertainties are estimated using the 
{\tt MSTW2008NNLO} PDF set. Uncertainties of 100\% are assumed for the multi-jet prediction and for the cross-section of \ttbar\ and $W$/$Z$ 
production in association with a $b\bar{b}$ pair.  For $\ttbar$+$W$/$Z$ production,  an uncertainty of approximatively 70\%  has been derived 
from the variations of the factorisation and renormalisation scales and from the PDF uncertainties~\cite{dileptonSS}. 

The $\ttbar$+jets yield in each signal region is extrapolated from the measured number of events in the corresponding control region (as per 
Table~\ref{tab:CR}) using a fit based on the profile likelihood method~\cite{profilelike}.  Each pair of control and signal regions is fitted 
separately, assuming no signal events. The free parameter in each fit is the $\ttbar$+jets overall normalisation scale, while the contributions 
from subdominant background processes are fixed at the expected value. Systematic uncertainties are treated as nuisance parameters constrained 
with a Gaussian function and correlations are taken into account where appropriate. The results of the fits and the numbers of observed events 
for each signal region are summarised  in Table~\ref{tab:results}. The fitted values of the normalisation factors for $\ttbar$+jets are compatible 
with one and the main impact of the data-driven estimate is a reduction in the uncertainty by approximately a factor of two. Figure~\ref{fig:SR} shows 
the measured \meff\ distributions and the MC predictions for the SM backgrounds in each signal region. Also shown are the prediction of two benchmark 
signal models described below. 

\begin{figure*}
\centering
\includegraphics[width=\mywidth\columnwidth]{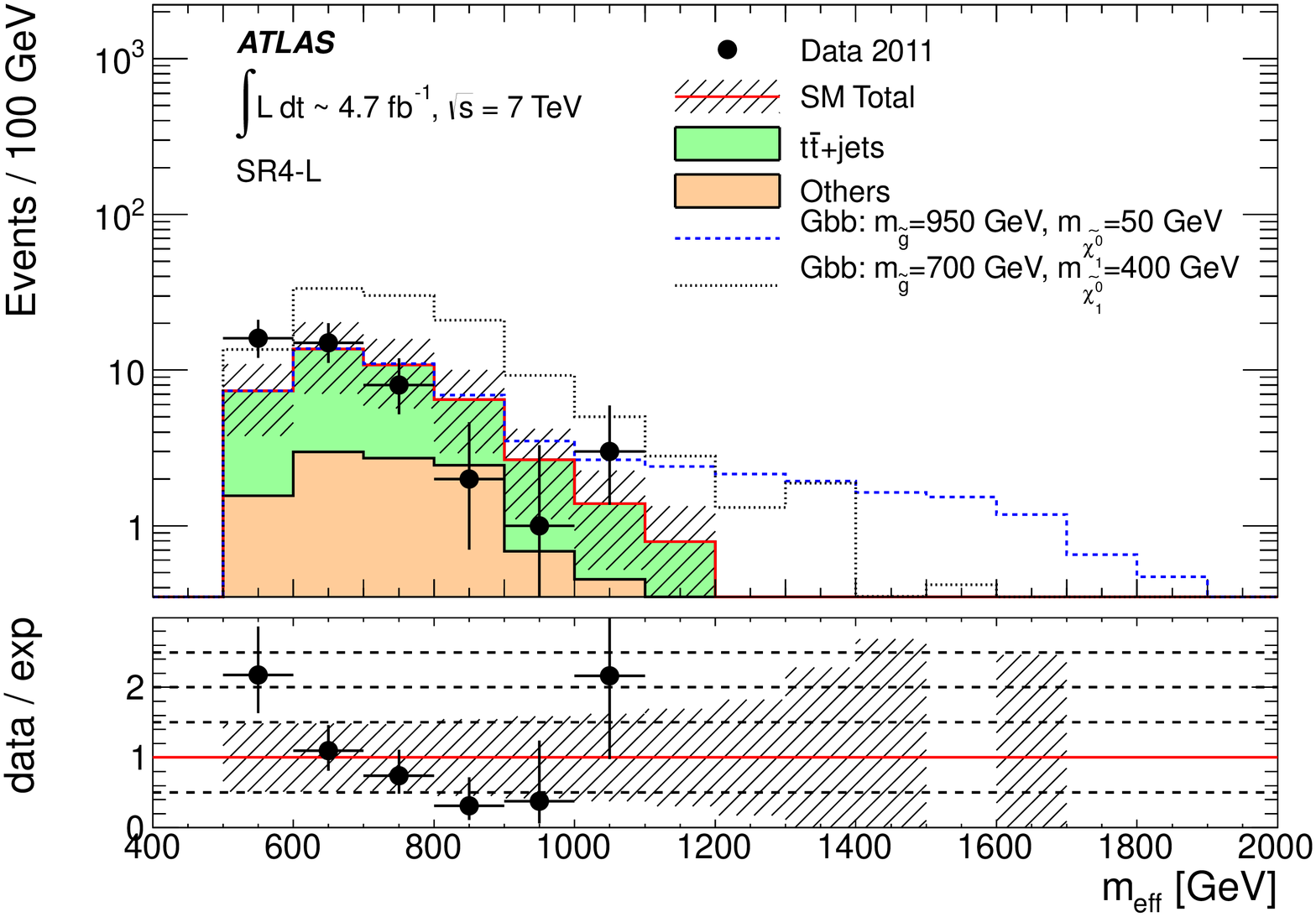}
\includegraphics[width=\mywidth\columnwidth]{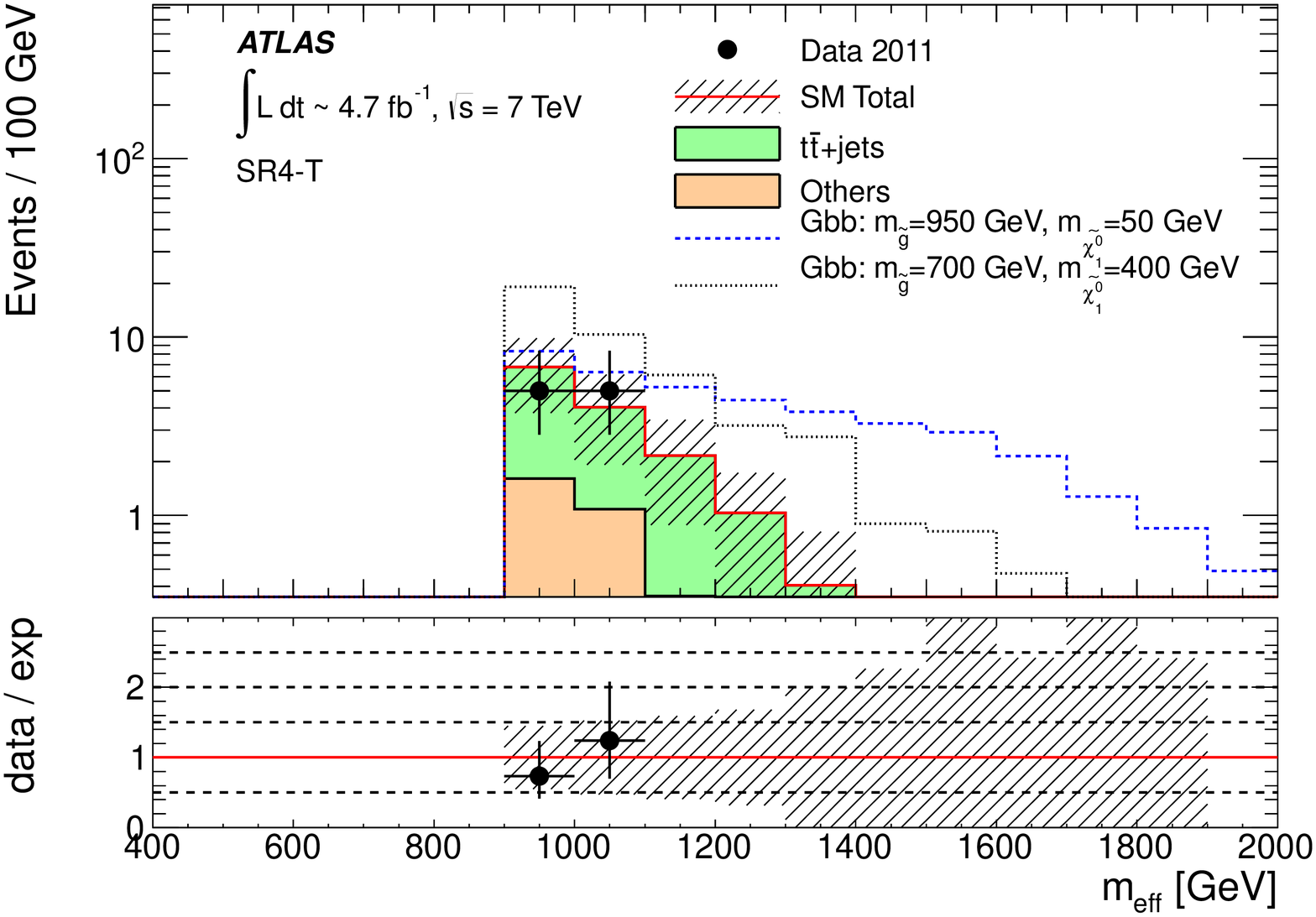} \\
\includegraphics[width=\mywidth\columnwidth]{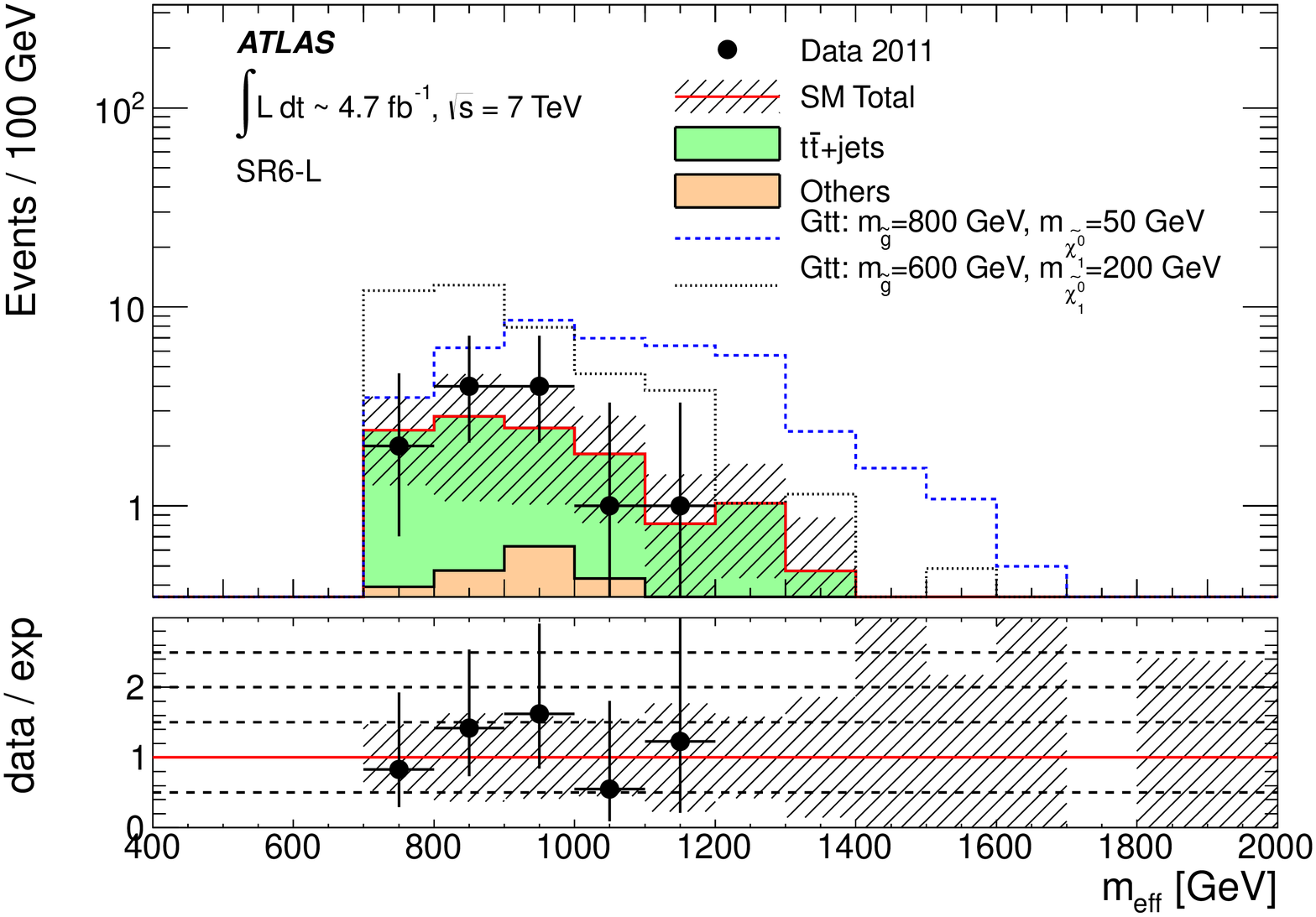}
\includegraphics[width=\mywidth\columnwidth]{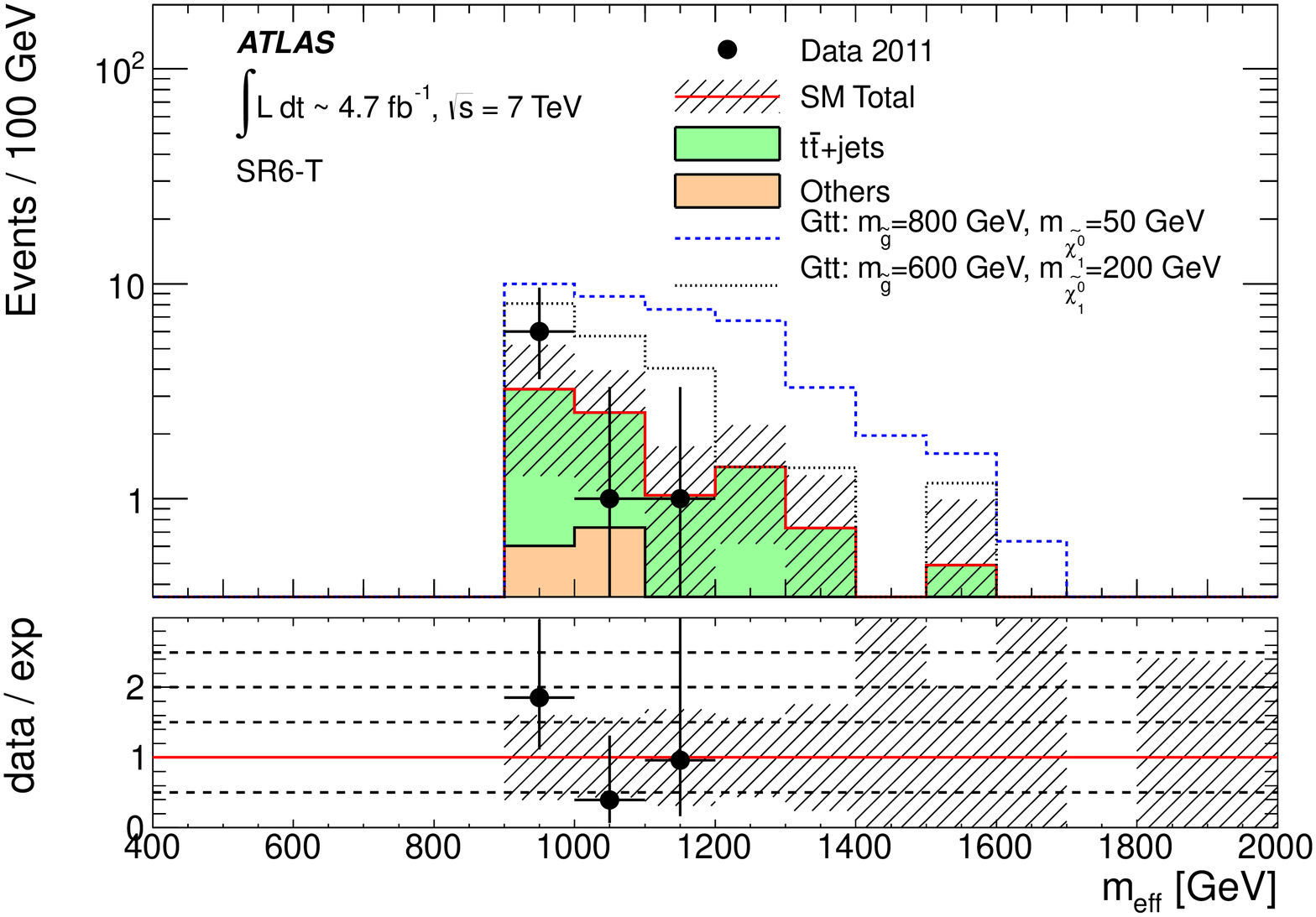}
\caption{Distribution of \meff\ for SR4-L and SR4-T (top) and SR6-L and SR6-T (bottom). The hatched band shows the systematic uncertainty on the 
MC prediction, which includes both experimental uncertainties (among which JES and $b$-tagging uncertainties are dominant) and theoretical 
uncertainties on the background normalisation and shape. The label ``others'' includes the contributions from single top, $\ttbar$+$b\bar{b}$, 
$\ttbar$+$W$/$Z$, $W$/$Z$+jets and multi-jet processes. The lower plot in each figure shows the ratio of the observed distribution to that expected 
for the SM background. Two signal points (with small and large mass splitting between the gluino and the LSP) for the \emph{Gbb} and 
\emph{Gtt} models described in the text are overlaid.} 
\label{fig:SR}
\end{figure*}

\begin{table}[htp]
\centering
\begin{tabular}{c|c|c|c|c}
\hline
\multirow{2}{*}{SR} & $\ttbar$+jets & others & SM  &  data \\
                    & (MC)    &        &        &  \\
\hline
\multirow{2}{*}{\SRA}& 33.3 $\pm$ 7.9  & 11.1 $\pm$ 4.9 & 44.4 $\pm$ 10.0  & 45 \\
                     & (32.6 $\pm$ 15.4)     &                  &                 &  \\
\multirow{2}{*}{\SRB} & 16.4 $\pm$ 4.1 & 6.6 $\pm$ 2.9 & 23.0 $\pm$ 5.4 & 14 \\ 
                     & (16.1 $\pm$ 8.4)     &                  &                 &  \\
\multirow{2}{*}{\SRC} & 9.6 $\pm$ 2.1 & 3.7 $\pm$ 1.6 & 13.3 $\pm$ 2.6  &  10 \\ 
                     & (11.4 $\pm$ 5.4)     &                  &                 &  \\
\hline
\multirow{2}{*}{\SRD} & 10.3 $\pm$ 3.3 & 2.4 $\pm$ 1.4 &  12.7 $\pm$ 3.6  &  12 \\ 
                     & (10.0 $\pm$ 6.2)     &                  &                 &  \\
\multirow{2}{*}{\SRE} & 8.3 $\pm$ 2.4  & 1.6 $\pm$ 1.1 & 9.9 $\pm$ 2.6 &  8 \\ 
                     & (7.9 $\pm$ 5.3)     &                  &                 &  \\
\hline
\end{tabular}
\caption{Comparison between the results of the fits and the numbers of observed events in the five signal regions. The $\ttbar$+jets event yield predicted 
by the MC simulation is quoted in parentheses. The column ``others'' includes the contributions from single top,  $\ttbar$+$b\bar{b}$, $\ttbar$+$W$/$Z$, 
$W$/$Z$+jets and multi-jet processes. Multi-jet events contribute less than 5\% of the total background. The column ``SM'' shows the total expected 
background and is the sum of the columns ``$\ttbar$+jets'' and ``others''. The uncertainties include the statistical and systematic uncertainties.} 
\label{tab:results}
\end{table}

The reliability of the MC extrapolation of the \ttbar\ background to larger $b$-jet multiplicities has been checked in validation regions defined 
with kinematic cuts similar to those used in the control and signal regions, except that exactly one isolated electron or muon is required. The 
transverse mass of the lepton and the \met\ is required to be less than 100~\gev\ in all validation regions to minimise the possible contamination 
from stop production. The extrapolated event yield in the validation regions with at least three $b$-jets  from the validation 
regions with exactly two $b$-jets is found to be consistent with the number of observed events for all $b$-tagging operating points.

The background predictions have been further validated using a data-driven method that simultaneously  estimates all SM background contributions with 
at least one misidentified $b$-jet. This method consists of predicting the number of jets originating from $b$-quarks in each event by solving a system 
of equations based on the number of $b$-tagged and non $b$-tagged jets in the event, along with  the $b$-tagging efficiency and mistag rates. 
Consistent background predictions with respect to the fit results have been found in all signal regions. 

Limits for non-SM signal at 95\% confidence level (CL) are derived by testing the signal plus background hypothesis in each signal region with the 
$CL_s$ prescription~\cite{profilelike}. These limits are obtained with fits similar to those used to estimate the background in each signal region, 
except that the number of observed events in the signal region is added as an input to the fit and a second free parameter for the non-SM signal 
strength, constrained to be non-negative, is adjusted in the likelihood maximisation. This additional free parameter ensures a proper treatment of 
the expected signal contamination in the control regions when the results are interpreted in the framework of specific SUSY scenarios. Model-independent 
upper limits at 95\% CL on the number of signal events and on the visible cross-section (defined as the cross-section times kinematic acceptance times 
experimental efficiency) for non-SM contributions derived for each signal region are given in Table~\ref{tab:UL}. 

\begin{table}[htp]
\centering
\begin{tabular}{c|>{\centering\arraybackslash} p{2.5cm}|>{\centering\arraybackslash} p{2.5cm}}
\hline 
   & \multicolumn{2}{|c}{Obs (exp) 95\% CL  upper limit} \\
\hline 
SR &  $N$ events & $\sigma_{\mathrm{vis}}$(fb)  \\ 
\hline
\SRA &    23.8 (23.4) &   5.1 (5.0) \\
\SRB &   8.6 (12.8) &   1.8 (2.7) \\
\SRC &   7.1 (9.2) &   1.5 (2.0) \\ \hline
\SRD &  9.6 (10.1)  &   2.0 (2.1) \\ 
\SRE &   7.1 (8.3) &   1.5 (1.8) \\
\hline
\end{tabular}
\caption{Observed (expected) 95\% CL upper limits on the non-SM contributions to all signal regions. Limits are given on numbers of events and in 
terms of visible cross-sections defined by cross-section times kinematic acceptance times experimental efficiency. 
Systematic uncertainties on the SM background estimation are included in the limits.}  
\label{tab:UL}
\end{table}

\begin{figure*}
\centering
\includegraphics[width=\mywidth\columnwidth]{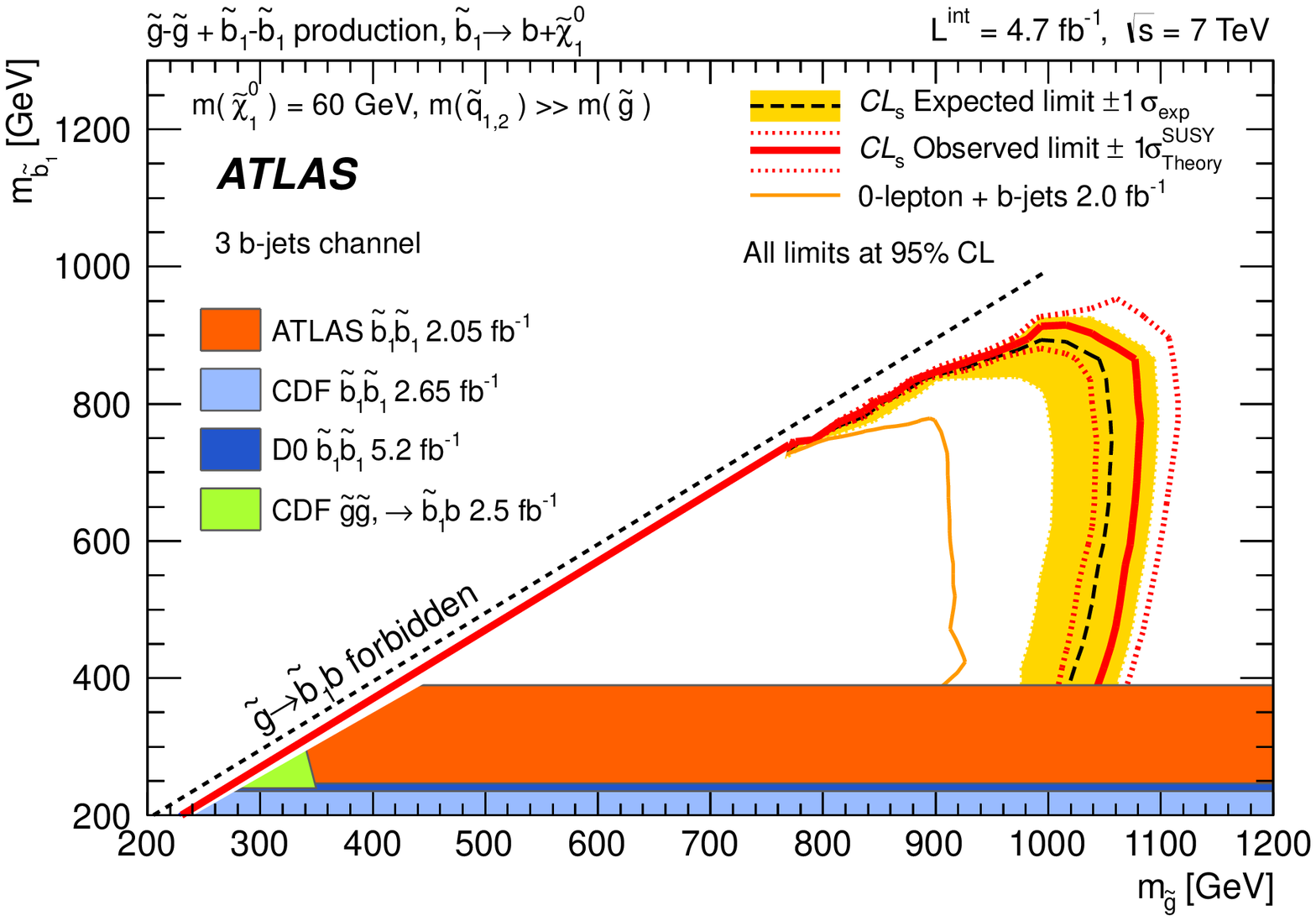}
\includegraphics[width=\mywidth\columnwidth]{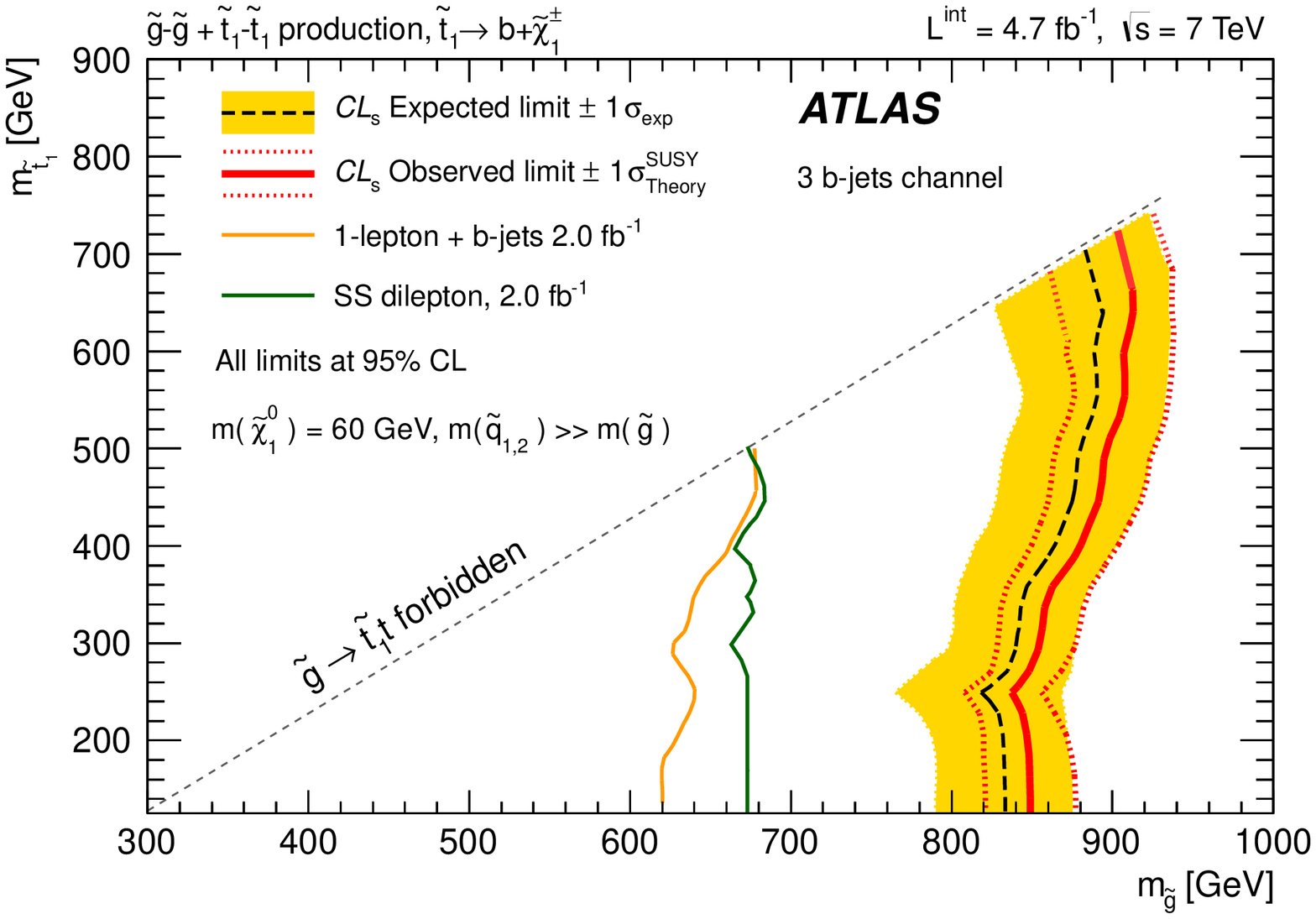} \\
\includegraphics[width=\mywidth\columnwidth]{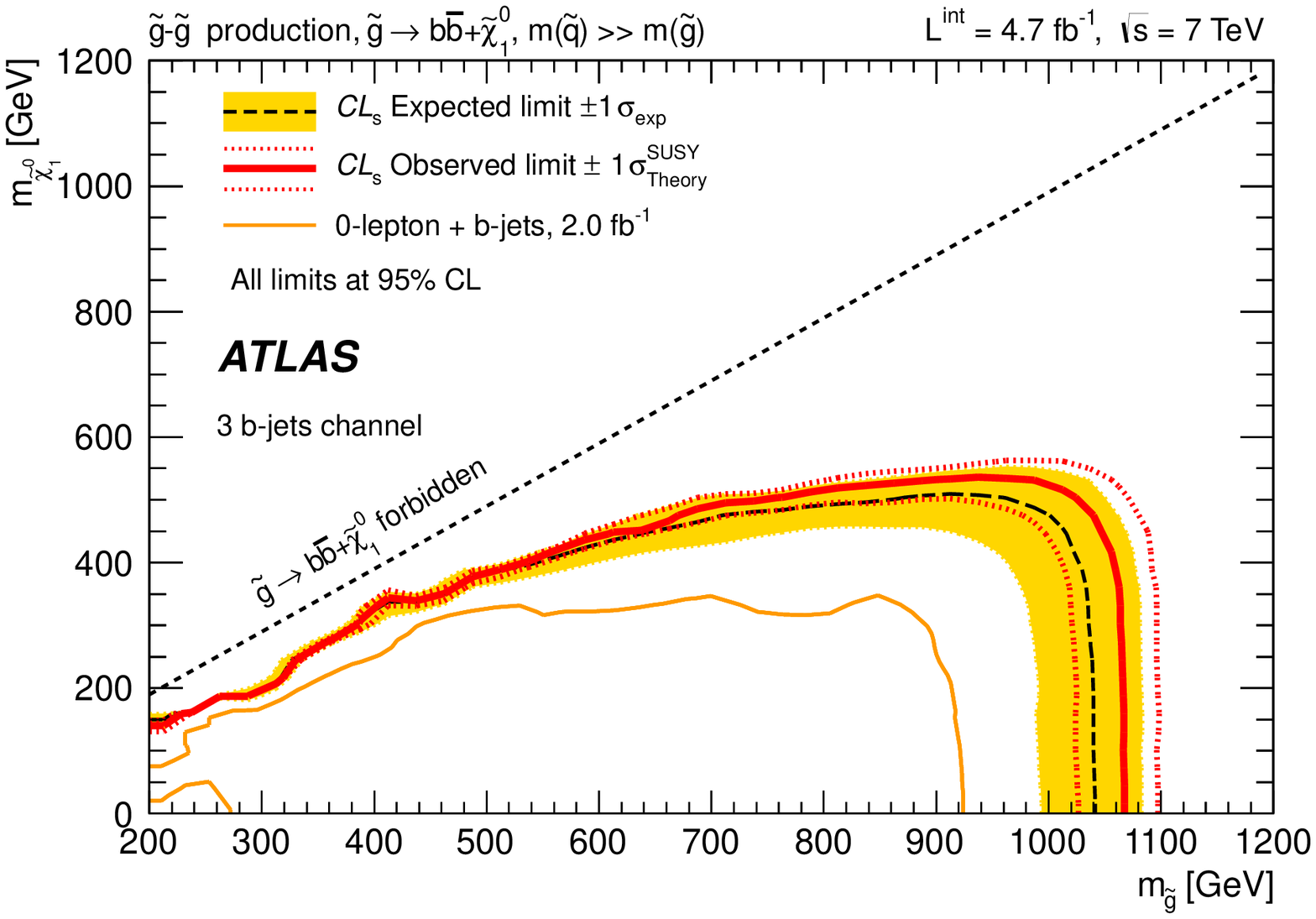}
\includegraphics[width=\mywidth\columnwidth]{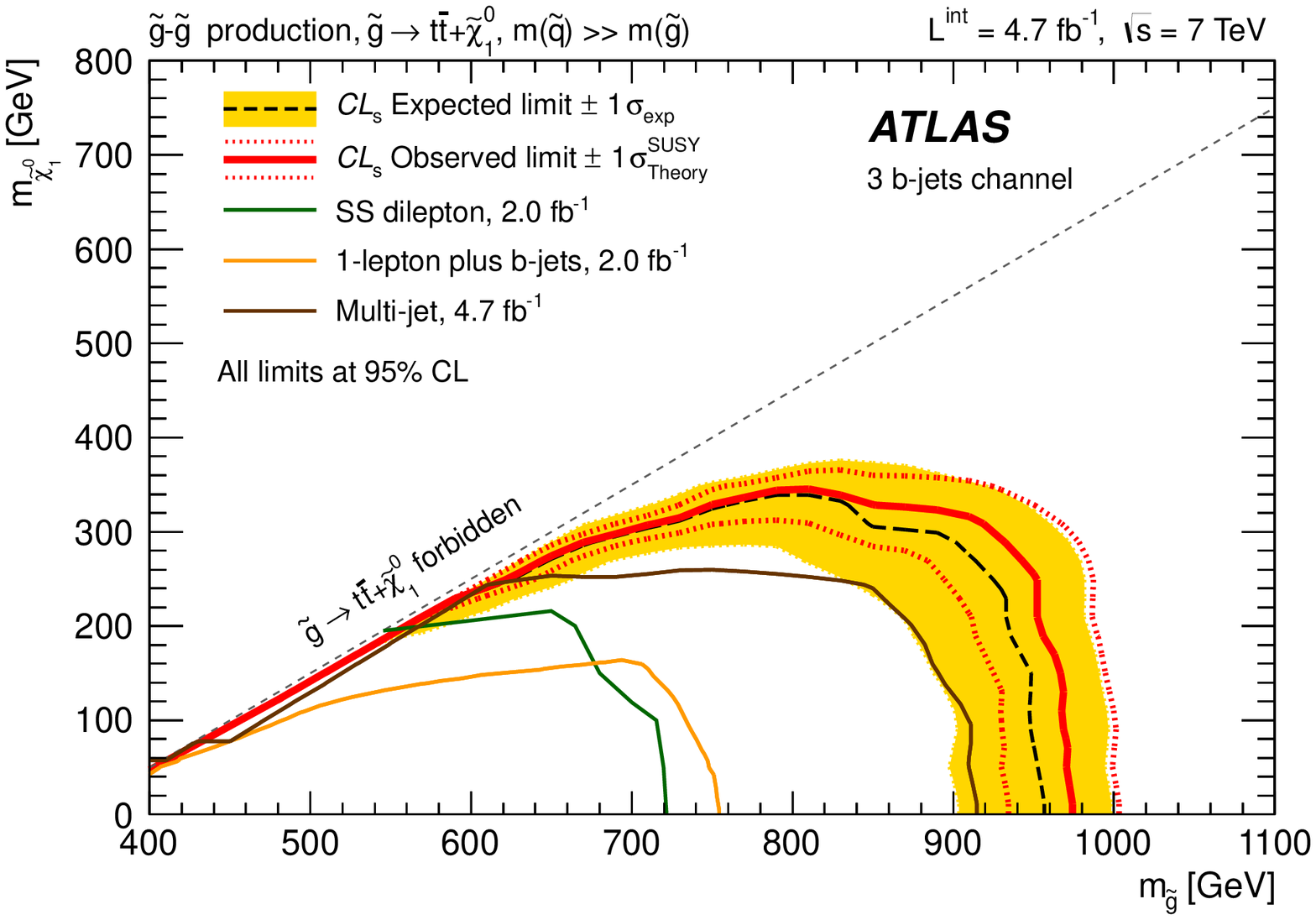} \\
\caption{Exclusion limits in the ($m^{}_{\gl},m^{}_{\bone}$) plane for the gluino-sbottom model (top left), in the ($m^{}_{\gl},m^{}_{\tone}$) plane 
for the  gluino-stop model (top right) and in the  ($m_{\tilde{g}},m_{\tilde{\chi}_1^0}$) plane for the \emph{Gbb} (bottom left) and \emph{Gtt} 
(bottom right) models. The dashed black and solid bold red lines show the 95\% CL expected and observed limits respectively, including all 
uncertainties except the theoretical signal cross-section uncertainty. The shaded (yellow) bands around the expected limits show the impact of the 
experimental uncertainties while the dotted red lines show the impact on the observed limit of the variation of the nominal signal cross-section by 
1$\sigma$ theoretical uncertainty. Also shown for reference are the previous CDF~\protect\cite{gluinosbottomCDF,sbottomCDF}, D0~\protect\cite{sbottomD0} and 
ATLAS~\protect\cite{sbottom,PRD,dileptonSS,multijet} analyses.}
\label{fig:limits}
\end{figure*}

These data have been used to derive limits in the parameter space of the following SUSY models: 

{\bf Gluino-sbottom model: } MSSM scenarios where the $\bone$ is the lightest squark, all other squarks are heavier than the gluino, and 
$m^{}_{\gl}>m^{}_{\bone}>m^{}_{\neut}$, so the branching ratio for $\gl \rightarrow \bone b$ decays is 100\%. Sbottoms are produced via $\gl\gl$ or 
by $\bone\bone$ direct pair production and are assumed to decay exclusively via $\bone \rightarrow b\neut$, where $m^{}_{\neut}$ is set to 
60~\GeV. Exclusion limits are presented in the ($m^{}_{\gl},m^{}_{\bone}$) plane.

{\bf \emph{Gbb} model: } Simplified scenarios, where $\bone$ is the lightest squark but $m^{}_{\gl}<m^{}_{\bone}$. Pair production of gluinos is 
the only process taken into account since the masses of all other sparticles apart from the $\neut$ are set above the \TeV\ scale. 
A three-body decay via an off-shell sbottom is assumed for the gluino, yielding a 100\% BR for the decay $\gl \rightarrow b\bar{b}\neut$. 
The sbottom mass has no impact on the kinematics of the decay and the exclusion limits are presented in the ($m^{}_{\gl},m^{}_{\neut}$) plane.  

{\bf Gluino-stop model: } MSSM scenarios where the $\tone$ is the lightest squark, all other squarks are heavier than the gluino, and 
$m^{}_{\gl}>m^{}_{\tone}+m^{}_{t}$, so the branching ratio for $\gl \rightarrow \tone t$ decays is 100\%. Stops are produced via $\gl\gl$ and 
$\tone\tone$ and are assumed to decay exclusively via $\tone \rightarrow b \chipm$. The neutralino mass is set to 60~\GeV, the chargino mass 
to 120~\GeV\ and the latter is assumed  to decay through a virtual $W$ boson. Exclusion limits are presented in the ($m^{}_{\gl},m^{}_{\tone}$) plane. 

{\bf \emph{Gtt} model: }  Simplified scenarios, where $\tone$ is the lightest squark but $m^{}_{\gl}<m^{}_{\tone}$. Pair production of gluinos is 
the only process taken into account since the mass of all other sparticles apart from the $\neut$ are above the \TeV\ scale. A three-body decay via 
off-shell stop is assumed for the gluino, yielding a 100\% BR for the decay $\gl \rightarrow t\bar{t}\neut$. The stop mass has no impact on the 
kinematics of the decay and the exclusion limits are presented in the ($m^{}_{\gl},m^{}_{\neut}$) plane.  

The SR4 regions are mostly sensitive to the SUSY models where sbottom production dominates, whilst the SR6 regions are used to set exclusion 
limits in models characterised by on-shell or off-shell stop production, where top-enriched final states are expected. The signal region with 
the best expected sensitivity at each point in the parameter space is used to derive the limits at 95\% CL. Signal cross-sections are calculated 
to next-to-leading order in the strong coupling constant, including the resummation of soft gluon emission at next-to-leading-logarithmic accuracy 
(NLO+NLL)~\cite{Beenakker:1996ch,Kulesza:2008jb,Kulesza:2009kq,Beenakker:2009ha,Beenakker:2011fu}. The nominal cross-section and the uncertainty 
$\sigma^{\mathrm{SUSY}}_{\mathrm{Theory}}$  are taken from an envelope of cross-section predictions using different PDF sets and factorisation and 
renormalisation scales, as described in Ref.~\cite{Kramer:2012bx}. All detector-related systematic uncertainties are treated as fully correlated 
between signal and backgrounds. In the \emph{Gbb} scenario, the impact of initial-state radiation (ISR) is expected to be large in the region with 
low $m_{\gl} - m_{\neut}$ due to the small signal acceptance. Therefore, an uncertainty on the modelling of ISR is assessed by comparing the signal 
acceptance obtained with the {\tt Herwig++} samples to the one obtained with dedicated {\tt MADGRAPH} samples generated with additional jets. 
This uncertainty varies from 4\% to 35\% as a function of $m_{\gl} - m_{\neut}$ and is included in the $\pm 1 \sigma^{\mathrm{SUSY}}_{\mathrm{Theory}}$ band. 

The expected and observed 95\% CL exclusion limits in the four models considered above are shown in Figure~\ref{fig:limits}. 
In the gluino-sbottom model, gluino masses below 1000~\GeV\ are excluded for sbottom masses up to about 870~\GeV\ using the most conservative 
$-1 \sigma^{\mathrm{SUSY}}_{\mathrm{Theory}}$ hypothesis. This extends by approximatively 100~\gev\ the limits derived in the same scenario by 
the previous ATLAS analysis performed with 2~$\ifb$~\cite{PRD} and is complementary to the ATLAS search for direct sbottom pair production, 
also carried out with 2~$\ifb$~\cite{sbottom}. The exclusion is less stringent in the region with low $m_{\gl}-m_{\bone}$, where softer jets are 
expected. Because of the kinematic cuts applied, the limits depend on the neutralino mass assumption for low mass splitting between the sbottom and 
the neutralino as shown for the \emph{Gbb} model where gluino masses below 1020~\GeV\ are excluded for neutralino masses up to about 400~\GeV, 
improving the previous ATLAS limits~\cite{PRD} by approximatively 100~\GeV. In the gluino-stop model, gluino masses below 820~\GeV\ are excluded 
for stop masses up to 640~\GeV, extending the previous ATLAS limits~\cite{PRD,dileptonSS} by approximatively 150~\GeV. In the \emph{Gtt} model, 
gluino masses below 940~\GeV\ are excluded for $m_{\neut} < 50$~\GeV\ while neutralino masses  below 320~\GeV\ are excluded for $m_{\tilde{g}} = 800$~\GeV. 
This search extends the exclusion limits on the gluino mass from the ATLAS multi-jet analysis carried out with the same data set~\cite{multijet} and 
from the CMS same-sign dilepton analysis performed with 5~$\ifb$~\cite{cms_2llSS} by approximatively 60~\gev\ and 130~\GeV, respectively, for 
neutralino masses  below 100~\GeV. In the region with low $m_{\gl}-m_{\neut}$, the limits obtained with the CMS analysis are most stringent 
due to the softer kinematic cuts.  

In summary, this letter presents results from a search for top and bottom squarks in the decay of gluino pairs produced in $pp$ collisions at 
$\sqrt{s}$ = 7~\TeV, based on \lumi~$\ifb$ of ATLAS data. The events are selected with large \met, four or six jets and at least three jets originating from 
\bquarks in the final state. The results are in agreement with the SM background prediction and translate into 95\% CL upper limits on excluded masses 
for a variety of SUSY benchmark scenarios. Gluino masses up to 1~\TeV\ are excluded, depending on the model, which significantly extends the previous results. 

We thank CERN for the very successful operation of the LHC, as well as the support staff from our institutions without whom ATLAS could not be 
operated efficiently. We acknowledge the support of ANPCyT, Argentina; YerPhI, Armenia; ARC, Australia; BMWF, Austria; ANAS, Azerbaijan; SSTC, 
Belarus; CNPq and FAPESP, Brazil; NSERC, NRC and CFI, Canada; CERN; CONICYT, Chile; CAS, MOST and NSFC, China; COLCIENCIAS, Colombia; MSMT CR, 
MPO CR and VSC CR, Czech Republic; DNRF, DNSRC and Lundbeck Foundation, Denmark; ARTEMIS, European Union; IN2P3-CNRS, CEA-DSM/IRFU, France; GNAS, 
Georgia; BMBF, DFG, HGF, MPG and AvH Foundation, Germany; GSRT, Greece; ISF, MINERVA, GIF, DIP and Benoziyo Center, Israel; INFN, Italy; MEXT and 
JSPS, Japan; CNRST, Morocco; FOM and NWO, Netherlands; RCN, Norway; MNiSW, Poland; GRICES and FCT, Portugal; MERYS (MECTS), Romania; MES of Russia 
and ROSATOM, Russian Federation; JINR; MSTD, Serbia; MSSR, Slovakia; ARRS and MVZT, Slovenia; DST/NRF, South Africa; MICINN, Spain; SRC and 
Wallenberg Foundation, Sweden; SER, SNSF and Cantons of Bern and Geneva, Switzerland; NSC, Taiwan; TAEK, Turkey; STFC, the Royal Society and 
Leverhulme Trust, United Kingdom; DOE and NSF, United States of America. The crucial computing support from all WLCG partners is acknowledged 
gratefully, in particular from CERN and the ATLAS Tier-1 facilities at TRIUMF (Canada), NDGF (Denmark, Norway, Sweden), CC-IN2P3 (France), 
KIT/GridKA (Germany), INFN-CNAF (Italy), NL-T1 (Netherlands), PIC (Spain), ASGC (Taiwan), RAL (UK) and BNL (USA) and in the Tier-2 facilities 
worldwide.

\bibliographystyle{spphys}      
\bibliography{ATLAS_SUSY_3bjets}

\clearpage 
\onecolumn
\begin{flushleft}
{\Large The ATLAS Collaboration}

\bigskip

G.~Aad$^{\rm 47}$,
T.~Abajyan$^{\rm 20}$,
B.~Abbott$^{\rm 110}$,
J.~Abdallah$^{\rm 11}$,
S.~Abdel~Khalek$^{\rm 114}$,
A.A.~Abdelalim$^{\rm 48}$,
O.~Abdinov$^{\rm 10}$,
R.~Aben$^{\rm 104}$,
B.~Abi$^{\rm 111}$,
M.~Abolins$^{\rm 87}$
% ...
\bigskip

% ...
$^{10}$ Institute of Physics, Azerbaijan Academy of Sciences, Baku, Azerbaijan\\
$^{11}$ Institut de F\'{i}sica d'Altes Energies and Departament de F\'{i}sica de la Universitat Aut\`{o}noma de Barcelona and ICREA, Barcelona, Spain\\
% ...
$^{20}$ Physikalisches Institut, University of Bonn, Bonn, Germany\\
% ...
$^{47}$ Fakult\"{a}t f\"{u}r Mathematik und Physik, Albert-Ludwigs-Universit\"{a}t, Freiburg, Germany\\
$^{48}$ Section de Physique, Universit\'{e} de Gen\`{e}ve, Geneva, Switzerland\\
% ...
$^{87}$ Department of Physics and Astronomy, Michigan State University, East Lansing MI, United States of America\\
% ...
$^{104}$ Nikhef National Institute for Subatomic Physics and University of Amsterdam, Amsterdam, Netherlands\\
% ...
$^{110}$ Homer L. Dodge Department of Physics and Astronomy, University of Oklahoma, Norman OK, United States of America\\
$^{111}$ Department of Physics, Oklahoma State University, Stillwater OK, United States of America\\
% ...
$^{114}$ LAL, Universit\'{e} Paris-Sud and CNRS/IN2P3, Orsay, France\\
% ...
\end{flushleft}


\end{document}